\documentclass[pra,twocolumn,showpacs]{revtex4}

\usepackage{graphicx}
\usepackage{bm,color}

\usepackage{graphicx}
\usepackage{amsmath}
\usepackage{bm,color}
\usepackage{multirow}

\newcommand{\be}{\begin{eqnarray}}
\newcommand{\ee}{\end{eqnarray}}

\renewcommand{\theequation}{\arabic{equation}}

\begin{document}

\title{Extended flat-band, entanglement and topological properties in a Creutz ladder}
\date{\today}

\author{Yoshihito Kuno}
\affiliation{Department of Physics, University of Tsukuba, Tsukuba, Ibaraki 305-8571, Japan}

\begin{abstract}
In this work, we study the entanglement and topological properties of an extended flat-band Creutz ladder by considering a compacted localized state (CLS). 
Based on the CLS picture, we find a multiple flat-band extension from the conventional two flat-band Creutz ladder. 
A simple vertical inter-chain coupling leads to a four complete flat-band system 
and creates an additive $\pi$-flux pattern on the Creutz ladder.
Interestingly, the strong coupling induces a topological phase transition where the distribution of CLSs is modified: 
upper and lower flat-band CLSs are paired up. 
This pairing leads to the destruction of the CLS' entanglement and, hence, to a vanishing edge mode (i.e., the breakdown of non-trivial topological phase).  
Finally, we study the localization dynamics induced by the presence of complete flat bands in this extended flat-band system.
\end{abstract}


\maketitle
\section{Introduction}
The investigation of flat-band models is currently one of the hot topics in condensed matter physics. 
So far, various theoretical models and artificially constructed flat-band systems have been founded and extensively studied \cite{Mielke,Tasaki1,vidal1,vidal2,Leykam0,Tang}.
In particular, concrete flat-band systems such as solid materials \cite{Hatsugai1,Hatsugai2,Mizoguchi1,Mizoguchi2}, 
synthetic photonic crystals \cite{Mukherijee1,Mukherijee2,Kremer,Vicencio0} and cold-atom optical lattice \cite{Taie,Taie2} have been actively researched. 
Also, in recent years, a metal-organic framework \cite{Liu0,Su,Jiang0} and kagome lattice \cite{Jiang3,Zhang0} have attracted attention from a flat band perspective. 
The rich physics of flat-band models generally originates from distractive hopping interferences. 
Flat-bands causes interesting phenomena: flat-band ferromagnetization \cite{Mielke,Tasaki1}, various quantum phases under interactions \cite{Takayoshi,Mondaini,Huber,Tovmasyan1,Tovmasyan2}, 
complete localization phenomena called Aharanov-Bohn (AB) caging \cite{vidal1,vidal2}, 
and fractional topological phenomena \cite{Bergholtz}. 
Recently, some artificial flat-band systems have been developed experimentally \cite{Leykam0}, 
also topologically flat-band models has been proposed theoretically \cite{Pal1,Pal2,Tang}.
In photonic crystal, so far a photonic Lieb lattice and a diamond flat-band lattice model have been realized 
and the AB caging dynamics was clearly simulated for both systems \cite{Mukherijee1,Mukherijee2,Kremer}. 
This is nothing but the observation of the localization of light induced by flat-band nature, 
and also the localization can be translated as disorder-free localization \cite{Flach,Flach2,Shukle1,Shukle2,Smith1,Smith2,Ours1,Zurita} and ergodicity breaking, 
which have been extensively discussed in the context of Anderson and many-body-localization studies \cite{Nandkishore,Abanin,Alet}.
Also, Lieb lattice, having one flat and two dispersive bands, 
has been realized in an optical lattice system in which the atoms were loaded in the flat-band state \cite{Taie,Taie2} 
and also real materials with Lieb lattice in organic frameworks was discovered \cite{Jiang}. 
The Creutz ladder \cite{Creutz} has been implemented by employing different orbital degrees of freedom of atom and using a lattice modulation technique \cite{Kang}. 

The Creutz ladder is distinguishes itself from the other flat-band models for
possessing two complete flat bands (at the condition of choosing suitable parameters). 
Therefore, the topological aspects of this particular model (with or without interactions) have been the subjects of extensive theoretical studies \cite{Bermudez1,Bermudez3,Bermudez4,Velasco}.
Additionally, the Creutz ladder is expected to exhibit exotic topological phases (e.g., the fractional topological phase) \cite{Barbarino}.
Overall, the Creutz ladder can be considered as a simple but interesting model, which exhibits both flat-band and topological features.

Motivated by both the most recent experimental achievements \cite{Kang,Mukherijee2} and previous theoretical studies on the Creutz ladder \cite{Bermudez2,Sun}, 
in this paper we pursue deeper understanding of the properties of the flat-band and topological phases that emerge on the Creutz ladder. 
The flat-band Creutz ladder can be described by a localized Wannier base called the compacted localized states (CLS). 
The CLS provides an intuitive picture of the AB caging and various quantum phases even in the case of interacting problems \cite{Takayoshi,Mondaini,Huber,Ours1,Zurita,Bermudez1,Bermudez3,Bermudez4,Tamura}. 
So far, in previous studies, the Creutz ladder has been recognized as two complete flat-band system. In this paper, we show an extension: from the CLS picture, multiple flat-band system can be constructed based on the conventional Creutz ladder. 
The extension is not so complicated: it requires only the addition of an inter-chain coupling and of an on-site potential, which lead to multiple flat-bands. 
The multiple flat-band Creutz ladder exhibits a new topological phase transition, clearly described by the distribution of the CLS. 
In particular the half-filled groundstate properties described by the CLS changes significantly through this new topological phase transition. 
We discuss entanglement and topological properties in detail with help of the CLS picture.  
Even in the multiple flat-band model, the AB caging is naturally preserved since the CLS system is a multiple band insulator.
This work shows rich flat-band topological phenomena based on the Creutz ladder system and also gives possibility of new topological phases some synthetic ladder experimental systems \cite{synthetic_dim1,synthetic_dim2}. 

The rest of this paper is organized as follows. 
In Sec. II, we introduce the Creutz ladder model. In Sec. III, we explain a recipe of the extension of flat-band. 
In Sec. IV, a phase transition is discussed in the extended flat-band system, 
and the properties of entanglement and topological phase are shown in Sec. V, and VI. 
In Sec. VII, we study a localization dynamics in the extended flat-band system.  We present our conclusions in Sec. VIII.

\section{Creutz ladder model}
\begin{figure}[t]
\begin{center} 
\includegraphics[width=8.5cm]{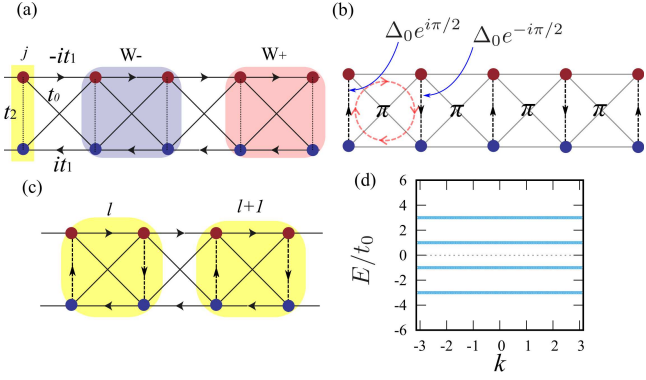}
\end{center} 
\caption{(a) A conventional Creutz ladder. (b) Additive $\pi$-flux induced by vertical inter-chain coupling. 
(c) Extended unit cell for the four-flat band system. (d) Bulk-band structure of $H_{\rm Cd}(k)$ for $\Delta_{0}=t_0$.}
\label{Setup}
\end{figure}

As a base model of this work, we start to consider the Creutz ladder originally introduced by Creutz \cite{Creutz}, 
\begin{eqnarray}
H_{\rm C}=&&\sum_{j}\biggl[-it_1(a^{\dagger}_{j+1}a_{j}-b^{\dagger}_{j+1}b_{j})\nonumber\\
&&-t_{0}(a^{\dagger}_{j+1}b_{j}+b^{\dagger}_{j+1}a_{j})-t_2 a^{\dagger}_{j}b_{j}
+\mbox{h.c.}\biggr],
\label{Creutz}
\end{eqnarray}
where  $a^{(\dagger)}_{j}$ and $b^{(\dagger)}_{j}$ are the fermion annihilation (creation) operators on the upper and lower chains, respectively. 
$j$ refers to a unit cell as shown in Fig.~\ref{Setup} (a). $t_1$, $t_{0}$ and $t_2$ are the intra-chain, inter-chain and vertical-uniform inter-chain hopping amplitudes, respectively. 
When $t_{0}=t_{1}$ and $t_2=0$, the model has two complete flat-bands with $E=\pm 2t_{0}$ due to the hopping interference. 
In what follows, we set $t_{0}=t_{1}$, $t_2=0$. 
The degenerate eigenstates of the upper and lower flat bands, are given by \cite{Takayoshi,Mondaini,Bermudez2}
\begin{eqnarray}
|W_{\pm,j}\rangle = W^{\dagger}_{\pm, j}|0\rangle 
= \frac{1}{2}\biggl[ia^{\dagger}_{j}+b^{\dagger}_{j}\mp a^\dagger_{j+1}
\mp ib^{\dagger}_{j+1}\biggr] |0\rangle,
\label{CLS}
\end{eqnarray}
where $|0\rangle$ is the empty state and $W^{(\dagger)}_{\pm}$ is a CLS annihilation (creation) operator \cite{Flach,Flach2,Vicencio0,Vicencio,Morales-Inostroza}. 
The CLS is a four-site complete localized state. 
Here, $\{W_{\alpha,j},W^{\dagger}_{\beta,k}\}=\delta_{\alpha\beta}\delta_{jk}$, i.e., the CLS can be regarded as a fermion particle.
By using $W^{\pm}_{j}$ operator, the flat-band Creutz ladder can be detangled as follows: 
\begin{eqnarray}
H_{\rm CLS}&=&\sum_{j}\biggl[-2t_0 W^{\dagger}_{-,j}W_{-,j}+2t_{0}W^{\dagger}_{+,j}W_{+,j}\biggl],
\label{FBH}
\end{eqnarray}
The CLSs can be regarded as two component resided particles on a lattice site $j$ \cite{Flach,Flach2}.
The CLS is a key element to extend the number of the flat-band.
\begin{figure*}[t]
\begin{center} 
\includegraphics[width=17.5cm]{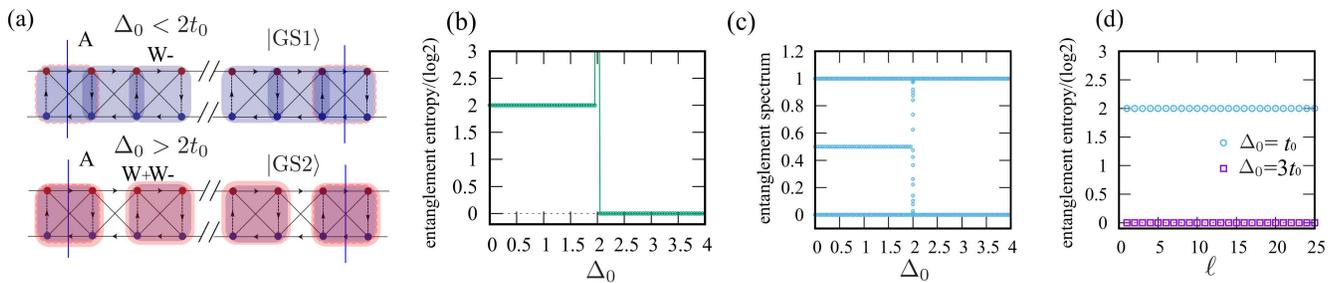}
\end{center} 
\caption{(a) Two types of groundstates described by the CLS. The blue vertical solid line represents the entanglement cut. 
Under a periodic boundary condition, the two lines divide the periodic system into two subsystem. 
(b) Entanglement entropy of the half-filled groundstate under varying $\Delta_{0}$. 
(c) Entanglement spectrum of the half-filled groundstate under varying $\Delta_{0}$. 
(d) Scaling behavior of the entanglement spectrum under varying subsystem size $\ell$.
In all cases, the ladder length is $L=50$, and $t_0=1$.}
\label{Fig2}
\end{figure*}

\section{Extension of flat-band}
The Hamiltonian $H_{\rm CLS}$ provides an insight of the extension from a two flat-band system to a multiple flat-band system.
If a staggered potential is added to $H_{\rm CLS}$, then  
\begin{eqnarray}
H_{\rm CLS2}=\sum_{j,s=\pm}\biggl[(2t_{s,0}+(-1)^{j}\Delta_{0})W^{\dagger}_{s,j}W_{s,j}\biggl],
\label{FBH2}
\end{eqnarray}
where $t_{s=\pm,0}=\pm t_{0}$ and $\Delta_{0}$ is the amplitude of the staggered potential. 
If $|\Delta_{0}|>0$, the two component CLS described by $H_{\rm CLS2}$ turns into a two band system 
since the system is described only by on-site terms. 
Hence, this model can be easily turned into a complete four-flat-band model. 
The Hamiltonian $H_{\rm CLS2}$ can be associated with the original Creutz ladder description by using the operators $a_{j}$ and $b_{j}$.
After some calculations (See appendix A), the Hamiltonian $H_{\rm CLS2}$ results equal to the following: 
\begin{eqnarray}
H_{\rm CLS2}=\left.H_{\rm C}\right|_{t_{0}=t_{1}}+\sum_{j}\biggl[(-1)^{j} i \Delta_{0} a^{\dagger}_{j}b_{j}+\mbox{h.c.}\biggr].
\label{HCV}
\end{eqnarray}
Interestingly, the staggered potential of $H_{\rm CLS2}$ gives the flat-band Creutz ladder only a simple vertical inter-chain coupling. 
Such CLS structure is preserved even if a vertical inter-chain coupling is added to the original Creutz ladder. 
This vertical inter-chain coupling assigns an additive $\pi$-flux to each plaquette of the ladder system (Fig.~\ref{Setup} (b)), and transforms the original two flat-band system into a four flat-band system.
Other multipleflat-band systems can be constructed as shown in appendix A. 

In addition, for a finite $\Delta_{0}$, the unit cell is extended to four sites (Fig.~\ref{Setup} (b)), the momentum representation of RHS of Eq.~(\ref{HCV}), here denoted as $H_{\rm Cd}$, can be given by 
\begin{eqnarray}
H_{\rm Cd}(k)=
\left[
\begin{array}{rrrr}
  0          & A          &     B(k) & C(k) \\
  A^{*}     & 0          & C(k) & D(k) \\
  B^{*}(k)  & C^{*}(k) &        0  & A^{*} \\
  C^{*}(k)      & D^{*}(k)  &       A  & 0 
\end{array}
\right],
\label{BMCd}
\end{eqnarray}
where $A = i\Delta_{0}$, $B(k)=it_{0}-it_{0}e^{-i k}$, $C(k)=-t_{0}-t_{0}e^{-ik}$ and $D(k)=-it_{0}+it_{0}e^{-ik}$.
$H_{\rm Cd}(k)$ certainly gives four complete flat-bands: $E_{1}=-2t_{0}-\Delta_{0}$, $E_{3}=-2t_{0}+\Delta_{0}$, 
$E_{2}=2t_{0}-\Delta_{0}$ and $E_{4}=2t_{0}+\Delta_{0}$. The band obtained by numerical calculation for $H_{\rm Cd}(k)$ is shown in Fig.~\ref{Setup} (d). 
In addition, we checked the effect of an arbitrary phase for the vertical hopping term in Eq.~(\ref{HCV}). 
The phase dependence of the band structure was numerically calculated to capture the transition of the band structure from two dispersive bands to four flat-bands. The results are shown in appendix B.
 
\section{Phase transition at half-filling}
Based on $H_{\rm CLS2}$, we can expect a direct phase transition under varying $\Delta_{0}$ at half-filling (i.e., when the two lowest bands are occupied).
For $\Delta_{0}=0$ and under periodic boundary condition, a groundstate with system size $L$ can be simply described by the lower band CLS $W^{\dagger}_{-,j}$ \cite{Bermudez2}, 
\begin{eqnarray}
|{\rm GS1}\rangle =\prod^{L}_{j}W^{\dagger}_{-,j}|0\rangle.
\label{GS1}
\end{eqnarray} 
The groundstate $|{\rm GS1}\rangle$ is maintained in $\Delta_{0}<2t_{0}$. 
However, at $\Delta_{0}=2t_{0}$ the energy of the state $W_{-,i\in {\rm even}}|0\rangle$ is comparable to that of the state $W_{+,i\in {\rm odd}}|0\rangle$. Then the second touches to the third one. This is a signal of a phase transition. We will later show that this is a topological phase transition. 
Furthermore, for $\Delta_{0}>2t_{0}$, a groudstate described by the CLS appears: 
\begin{eqnarray}
|{\rm GS2}\rangle =\prod_{j\in {\rm odd}}W^{\dagger}_{+,j}W^{\dagger}_{-,j}|0\rangle.
\label{GS2}
\end{eqnarray}
Here, the upper and lower CLSs are located on the same odd site ($j\in {\rm odd}$) in the Creutz ladder picture: a paring of the upper and lower CLSs is created (Fig.~\ref{Fig2} (a)).
Therefore, the $\Delta_{0}$ increase at half-filling triggers a phase transition, where the CLS configuration changes considerably. In other words, the vertical inter-chain coupling induces a phase transition in the Creutz ladder, {\it while maintaining the flat-band structure and the CLS picture}. 
In this sense, the transition described here is essentially different from that reported in a previous study \cite{Sun}. 

\section{Entanglement in the four flat-band system}
The CLS picture indicates that the vertical inter-chain coupling increases the number of flat bands in the system from two to four, and that the coupling induces a phase transition when  $\Delta_{0}=2t_{0}$. 
The entanglement and edge mode properties of the conventional flat-band Creutz ladder of $\left.H_{\rm C}\right|_{t_{0}=t_1}$, \cite{Bermudez2,Creutz}, as well as the effects of a vertical inter-chain coupling (i.e., a constant hopping term) in a non-flat ladder model have been previously discussed \cite{Nehra,Nehra2}.  
We examined the entanglement properties by considering a finite $\Delta_{0}$ and employing Peschel's method \cite{Peschel,Peschel2,Ryu,Mondragon1,Mondragon2}. We calculated the entanglement entropy (EE) and the entanglement spectrum (ES) under a certain entanglement cut. 
Our calculation was carried out in a half-filled system with ladder length $L$ and under a periodic boundary condition. The system was cut into two halves: each subsystem had a ladder length equal to $L/2$. 
In such flat-band system, we expect that the basic contribution unit of the EE essentially comes from the cut of a single CLS. 
The vertical cut of a single CLS gives a basic contribution of the EE, then the basic contribution is $\log 2$. The simple calculation is shown in appendix C. 

Figures~\ref{Fig2} (b) and (c) show the dependence of the EE and ES on $\Delta_{0}$.
A clear topological phase transition at $\Delta_{0}=2t_{0}$ is observed. The value of the EE changes from $2\log 2$ to zero when $\Delta_{0}=2t_0$. 
For $\Delta_0<2t_0$, the groundstate is $|{\rm GS1}\rangle$, 
then the cuts on both sides of edge correspond to the vertical cuts of the two lower band CLSs located on the two edges of the subsystem.
In this situation, EE is equal to $2\log 2$. 
For $\Delta_{0}>2t_0$, the groundstate is $|{\rm GS2}\rangle$ and the EE is equal to zero. 
Cutting a single pairing of the lower and upper CLSs located on both edges of the system seems to contribute to its EE; however, this same action does not contribute to the EE of the subsystem. 
This fact can be understood by considering a single-pairing state on a single plaquette as shown in appendix C. 
The pairing state exhibits no tendency to create entanglements, although each CLS has a finite value of EE. The pairing seems to exhibit a kind of the distractive effect of the EE in the system. 
The calculation of the ES also behaves as same with the EE. 
For $\Delta_0<2t_0$, two maximally entangled modes (MEM) (for which ES = 1/2) appear \cite{Pouranvari}. This indicates nothing but the presence of edge modes in the subsystem \cite{Ryu}.
For $\Delta_{0}>2t_0$, instead, the value of ES is always $0$ or $1$. 
This indicates the presence of two complete flat-bands without the MEM.
From this behavior, the phase transition at $\Delta_0=2t_0$ can be expected to be a topological one.
In addition, we calculated the scaling behavior of the EE by considering a fixed ladder length $L$ and varying the subsystem size $\ell$. For both phases($\Delta_0=t_0$ and $3t_0$), the EE exhibits a complete area law scaling (EE = const.).
This result confirms the flat-band nature of the system. 
At the critical point $\Delta_0=2t_0$, we found the EE behaves like a volume law scaling (EE$\propto$ $\ell$; data not shown).
Overall, the EE and ES results suggest that the $\Delta_{0}$ term considerably affects the entanglement properties of the system, matching the CLS picture. 
Next, we describe in detail the bulk topological properties.
\begin{figure}[t]
\begin{center} 
\includegraphics[width=8.5cm]{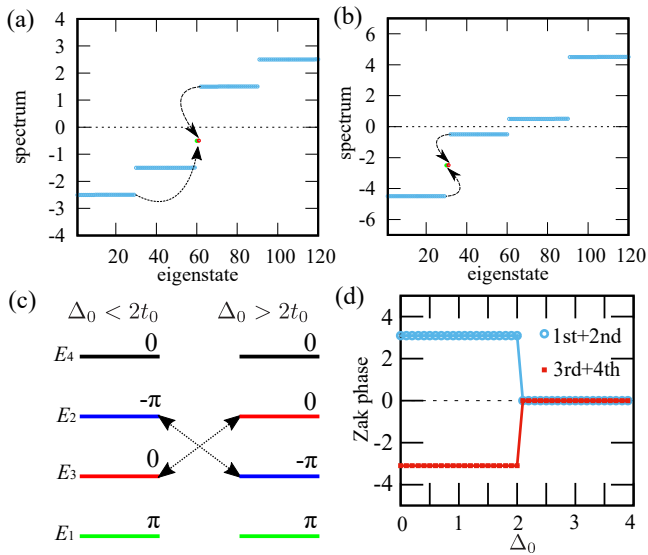}
\end{center} 
\caption{Energy spectrum for open boundary condition: (a) $\Delta_{0}<2t_{0}$, (b) $\Delta_{0}>2t_{0}$. 
The red and green dots represents the left and right localized edge modes. These modes have non-zero energy. 
(c) Band crossing of the second and third bands triggers a topological phase transition at half-filling. Each bulk band is characterized by Zak phase.
(d) Zak phase for lower-half and upper-half states. }
\label{Fig3}
\end{figure}

\section{Topological aspects of the flat-bands}
We characterize the bulk topological properties for a typical $\Delta_{0}$. 
The energy-spectrum for the open boundary system is plotted in Fig.~\ref{Fig3}.
Depending on the value of $\Delta_{0}$ the spectrum has edge modes between energy bands. 
For $\Delta_{0}=0.5 t_0 (<2t_{0})$, as shown in Fig.~\ref{Fig3} (a), two edge modes appear between the second and third bands.
These edge modes are given by a left localized mode $|{\rm L}\rangle = (1/\sqrt{2})[ a^{\dagger}_{1}+ib_1]|0\rangle$, 
and a right localized mode $|{\rm R}\rangle = (1/\sqrt{2})[ a^{\dagger}_{L}-ib_L]|0\rangle$, 
both of which correspond to the edge modes for the usual Creutz ladder \cite{Creutz,Bermudez2}, and have a finite energy proportional to $\Delta_0$, $H_{\rm CLS2}|{\rm L}({\rm R})\rangle =-\Delta_0|{\rm L}({\rm R})\rangle$. 
Notably, the edge mode $|{\rm L}({\rm R})\rangle$ comes from the first (third) bulk-band: 
if the system is at half-filling, only one edge mode is occupied. 
In the case of $\Delta_{0}>2t_{0}$ case, we find that the same two edge modes $|{\rm L}({\rm R})\rangle$ appear 
between the first and second band as shown in Fig.~\ref{Fig3} (b). 
Actually, the relative position between the edge modes and the flat band energy is clearly defined: 
for any value of $\Delta_0$($>0$), the edge modes appear just at the middle value between $E_1$ and $E_2$ flat bands i.e., $(E_1+E_2)/2=-\Delta_0$.
The edge mode $|{\rm L}({\rm R})\rangle$ comes from the first (second) bulk-band: if the system is at half-filling, both edge modes are occupied. 
Thus, the half-filled state in $\Delta_{0}>2t_{0}$ are essentially different from that in $\Delta_{0}<2t_{0}$.

By calculating the Zak phase we characterize the bulk topological index. 
We calculated the gauge invariant version of the Zak phase \cite{Vanderbilt} obtained from the real space eigenfunctions under a twisted boundary condition \cite{Barbarino}; more details are given in appendix D. The Zak phases of each flat-bands are shown in Fig.~\ref{Fig3} (c). 
The energy flat-bands $E_{1}$ and $E_{2}$ possess the non-trivial Zak phases $\phi_{Zak}=\pi$ and $-\pi$, respectively.
Actually, here the bulk-edge correspondence is confirmed since the energy flat-bands $E_{1}$ and $E_{2}$ generate edge modes under open boundary condition as shown in Fig.~\ref{Fig3} (a) and (b).
This result implies that the quarter-filled state is a non-trivial topological phase. 
Notably, when the system is at half-filling, the band crossing between the $E_{2}$ and $E_{3}$ energy flat-bands under varying $\Delta_0$ leads to a topological phase transition.  
We plot the Zak phase for the half-filled case in Fig.~\ref{Fig3} (d): the Zak phase for the lower-half (first + second band) states exhibits a clear topological phase transition at $\Delta_0=2t_0$. 
For the upper-half (third + fourth band) states, the Zak phase behaves in the opposite way. 
According to this bulk-edge topological character, the $\Delta_{0}$ term in Eq.~(\ref{HCV}) induces to a topological phase transition. 

\section{Localization dynamics}
\begin{figure}[t]
\begin{center} 
\includegraphics[width=7cm]{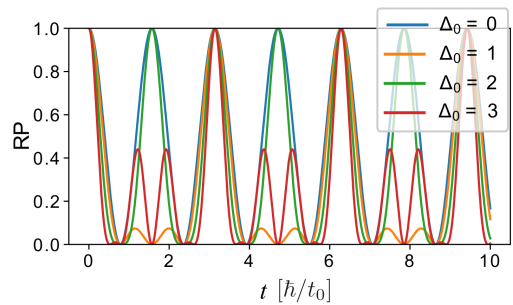}
\end{center} 
\caption{Return probability in a single particle dynamics. 
For any value of $\Delta_0$, the RP does not decay. $t_{0}=1$.}
\label{Fig4}
\end{figure}
The flat-band system described above also shows a localization tendency. 
In the case of the conventional flat-band Creutz ladder ($\Delta_0=0$), 
if we put on a single particles on a lattice site, the AB caging appears 
the particle is localized with oscillating and 
its exact dynamical solution has been known \cite{Zurita, Ours1}. 
The dynamics involved in the AB caging dynamics have been directly observed in recent experiments \cite{Mukherijee1,Mukherijee2,Kremer}. 
In the case of a finite $\Delta_0$, the localization dynamics are expected to be different. 
Therefore, we calculated the AB caging dynamics of a single particle. 
For the initial state, we consider a single localized particle, $|\Psi(0)\rangle = a^{\dagger}_{j}|0\rangle$.
We calculated the return probability (RP) to characterize the particle localization tendency, ${\rm RP}(t)=|\langle \Psi(0)|\Psi(t)\rangle |^2$, 
where $|\Psi(t)\rangle =e^{-iH_{\rm CLS2}t}|\Psi(0)\rangle$. 
The results for typical values of $\Delta_0$ are shown in Fig.~\ref{Fig4}. 
The localization is preserved Even for a finite $\Delta_0$ although the RP oscillation pattern is modified compared to the $\Delta_0=0$ case, where ${\rm RP}(t)=\cos^{2}(2t)$ \cite{Zurita,Ours1}, (unit of time is $\hbar/t_0$). 
Even for a finite $\Delta_0$, the particle dynamics are limited between $j-1$ and $j+1$ rungs. 
Accordingly, the AB caging is preserved and the localization length does not change even for the four complete flat-band model, and we expect that the four flat-band system does not thermalize even for many particle system prepared as a non-entangled state.

\section{Conclusion}
We have shown an extension of the conventional flat-band Creutz ladder. 
The recipe to construct a complete four flat-band model was shown with the help of the CLS picture.
Essentially, from the CLS picture, by extending this recipe, more multiple flat-band model can be constructed based on the original flat-band Creutz ladder. 

For the experimental realization of the extended four-flat band system, 
the implementation of the vertical inter-chain coupling in Eq.~(\ref{HCV}) is needed. This is challenging for present experimental systems, 
but, for example, Floquet photonic crystal systems \cite{Mukherijee2,Kremer} and 
cold-atom optical lattice systems with synthetic dimension and synthetic gauge fields \cite{synthetic_dim1,synthetic_dim2} 
have possibility to implement the vertical inter-chain coupling in Eq.~(\ref{HCV}).

The flat-band extension shown in this work might be applied to other flat-band models such as a diamond lattice.
Moreover, we found that the four complete flat-band exhibits a non-trivial topological phase transition, where the CLS distribution changes, 
and that the localization properties are preserved even in the four flat-band model.
Based on this four flat-band model, the detection of a fractional topological insulator \cite{Bergholtz} would be an interesting topic of study.  


\section*{Acknowledgments}
This work is supported by the Grant-in-Aid for JSPS Fellows (No.17J00486) and the JSPS KAKENHI (No. JP17H06138).

\renewcommand{\thesection}{A\arabic{section}} 
\renewcommand{\theequation}{A\arabic{equation}}
\renewcommand{\thefigure}{A\arabic{figure}}
\setcounter{equation}{0}
\setcounter{figure}{0}
\section*{Appendix A: CLS picture and introduction of a modulated potential}
In order to extend the two flat-band model to a four flat-band model and understand localization properties, 
it is useful to introduce the $w$-particle picture and the CLS picture.

First, we start with introducing the following operators:
\begin{eqnarray}
w^{\dagger}_{Aj}=ia^{\dagger}_j+b^{\dagger}_j,  \;\; w^{\dagger}_{Bj}=-ia^{\dagger}_j+b^{\dagger}_j,
\label{ww}
\end{eqnarray}
where we can prove $\{ w^\dagger_{Aj}, w_{Bj} \}=0$. We regard $w^{\dagger}_{A(B)j}$ as a particle, i.e., $w$-particle.
This transformation is a kind of detangling for a lattice system \cite{Flach}, 
then the $w$-particle representation of the flat-band Creutz ladder $\left.H_{\rm C}\right|_{t_{0}=t_{1}}$ is given by 
\begin{eqnarray}
\left.H_{\rm C}\right|_{t_{0}=t_{1}}=\sum_j\Big[it_0w^\dagger_{A,j}w_{Bj+1}-it_0w^\dagger_{Bj+1}w_{A,j}\Big].
\label{wparticle}
\end{eqnarray}
The above equation indicates that the A and B $w$-particles hopped to adjacent sites while the component changed from A to B or vice versa. 
However, these particles are prevented from extending into all lattice sites. In fact, the $\{ w_A, w_B\}$-particles are strictly localized on two adjacent rungs of the ladder. 
Furthermore, the $w$-particle operators are connected to the CLS operators of Eq.~(2) in the main text:  
\begin{eqnarray}
W^{\dagger}_{+,j}=\frac{1}{2}(w^{\dagger}_{Aj}-i w^{\dagger}_{Bj+1}), \;\; W^{\dagger}_{-,j}=\frac{1}{2}(w^{\dagger}_{Aj}+i w^{\dagger}_{Bj+1}).\nonumber\\
\label{WW_ww}
\end{eqnarray}
From these relations, we obtain $H_{\rm CLS}$ of Eq.~(3) in the main text. 
Let us consider introducing the staggered potential to $H_{\rm CLS}$. 
Then, the $w$-particle representation can be applied to $H_{\rm CLS2}$ without any difficulties, 
the representation is given by 
\begin{eqnarray}
H^{w}_{\rm CLS2}&&=\sum_j\Big[(it_0w^\dagger_{A,j}w_{Bj+1}-it_0w^\dagger_{Bj+1}w_{A,j}) \nonumber\\
&&+ (-1)^{j}\frac{\Delta_{0}}{2}(w^\dagger_{A,j}w_{A,j}+w^\dagger_{B,j+1}w_{Bj+1})\Big].\nonumber\\
\label{wparticle2}
\end{eqnarray}
The second term of the RHS is derived from the staggered potential term of $H_{\rm CLS2}$.
Through the $w$-particle representation, we obtain the second term in the RHS of Eq.~(5) in the main text. 
By using the relations shown in Eq.~(\ref{ww}), and shifting the summation index $j+1 \to j$ in the $w^\dagger_{B,j+1}w_{Bj+1}$ terms, 
the second term of the RHS in Eq.~(\ref{wparticle2}) can be represented by the original operators $a^{\dagger}_{j}$ and $b^{\dagger}_{j}$: 
\begin{eqnarray}  
&&(\mbox{Second term of Eq.~(\ref{wparticle2}}))\nonumber\\
&=&\sum_{j}\biggl[\frac{(-1)^{j}\Delta_{0}}{2}(w^\dagger_{A,j}w_{A,j}-w^\dagger_{B,j}w_{Bj})\biggr]\nonumber\\
&=&\sum_{j}\biggl[(-1)^{j}\Delta_{0} i a^{\dagger}_{j}b_{j}+\mbox{h.c.}\biggr].
\label{secondterm}
\end{eqnarray}
This is the second term in the RHS in Eq.~(5) in the main text.

By further extending the above method, we constructed a multiple-flat-band generalization based on the Creutz ladder. 
The following modulated potential of the CLS is added to $H_{\rm CLS}$ of Eq.~(3) in the main text:
\begin{eqnarray}  
V^{m}_n= \sum_{j,s=\pm}\biggl[\Delta_{0}\cos\biggl(\frac{2\pi j}{n}\biggr)W^{\dagger}_{s,j}W_{s,j}\biggl].
\label{n_flat_band_secondterm}
\end{eqnarray}
By adding this potential, it is possible to change the two-band insulator described by $H_{\rm CLS}$ into a $2n$-band insulator (considering the CLS system).
This change results in a $2n$ complete flat-band system $H_{\rm CLS}+V^{m}_{n}$. 
In the original Creutz ladder picture, the potential $V^{m}_n$ was given by 
\begin{eqnarray}  
V^{m}_n&=& \Delta_{0}\sum_{j}\biggl[\cos\biggl(\frac{2\pi j}{n}-\frac{\pi}{n}\biggr)\cos\biggl(\frac{\pi}{n}\biggr)(a^{\dagger}_{j}a_j+b^{\dagger}_{j}b_j)\nonumber\\
&&-\sin\biggl(\frac{2\pi j}{n}-\frac{\pi}{n}\biggr)\sin\biggl(\frac{\pi}{n}\biggr)(ia^{\dagger}_{j}b_j-ia_{j}b^{\dagger}_j)\biggl].
\label{n_flat_band_secondterm2}
\end{eqnarray}
In order to extend the original flat-band Creutz ladder to a $2n$ flat-band model, it is necessary to impose an on-site modulated potential along site $j$ and a complex vertical coupling in correspondence of site $j$.

\section*{Appendic B: Energy spectrum for arbitrary phase of the staggered vertical hopping}
\begin{figure}[h]
\begin{center} 
\includegraphics[width=8cm]{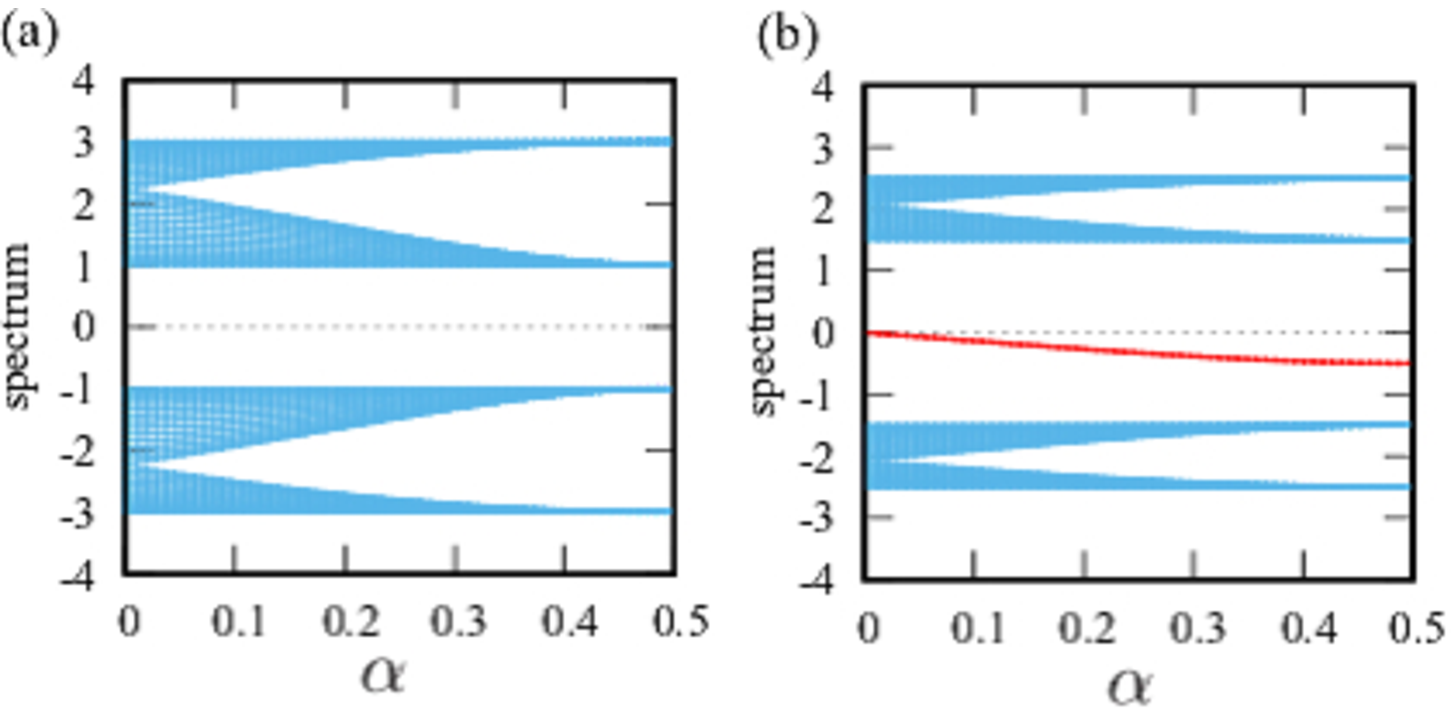}
\end{center} 
\caption{Energy spectrum: (a) Periodic boundary condition result with $\Delta_0=1$, 
(b) Open boundary condition result with $\Delta_0=0.5$. The red line is doubly degenerate, indicating the presence of both left and right localized edge modes. 
For both cases, $t_1=t_0=1$.}
\label{FigA0}
\end{figure}
To investigate the effect of the phase of the vertical coupling in Eq.~(5) in detail, we introduce an arbitrary phase of the staggered vertical hopping as follows:
\begin{eqnarray}
(-1)^j i \Delta_{0} a^{\dagger}_{j}b_{j}\to (-1)^{j}\Delta_{0}e^{-i\alpha \pi} a^{\dagger}_{j}b_{j}.
\label{alpha}
\end{eqnarray}
The parameter $\alpha$ induces a flux per a plaquette on the Creutz ladder.
We calculated the energy spectrum based on $H_{\rm CLS2}$ as varying $\alpha$ with $\Delta_0$ fixed. 
The results are displayed in Fig.~\ref{FigA0}(a) and (b).
Interestingly, for finite $\alpha$ the band splits from two to four, and also each band width decreases as $\alpha$ approaches to $0.5$, corresponding to Eq.(5) ($\pi$-flux pattern ).    
Furthermore, in Fig.~\ref{FigA0} (b), $\alpha$-dependence of the in-gap states is clearly captured. For $\alpha=0$, the in-gap state is doubly degenerate and zero-energy, corresponding to zero-energy edge states. For a finite $\alpha$, the in-gap states has a finite energy and for $\alpha=0.5$, the energy of the in-gap states becomes $-\Delta_0$.

\section*{Appendix C: Entanglement entropy for a single CLS on four site system}
\begin{figure}[h]
\begin{center} 
\includegraphics[width=4cm]{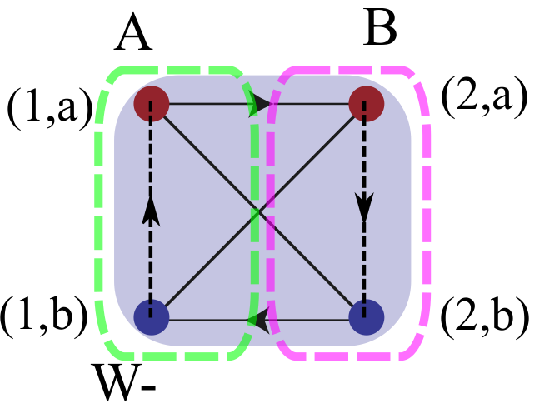}
\end{center} 
\caption{Single lower flat-band CLS}
\label{FigA1}
\end{figure}

In order to clarify the primitive entanglement properties of the CLS, 
we consider a simple four site system where a single lower CLS $W^{\dagger}_{-,1}|0\rangle$ is resided on a single plaquette, i.e., four sites as shown in Fig.~\ref{FigA1}.
Here, consider splitting the four-site system into two helves called A(B)-subsystem as shown in Fig.~\ref{FigA1}. 
The state of the single lower CLS is concretely expressed as follows:
\begin{eqnarray}
|W_{-,1}\rangle &=& \frac{1}{2}\biggl[ia^{\dagger}_{1}+ b^\dagger_{1}+a^{\dagger}_{2}
+ ib^{\dagger}_{2}\biggr] |0\rangle \nonumber\\
&=&\frac{1}{2}\biggl[|1,a\rangle_{A}|0\rangle_{B}+|1,b\rangle_{A}|0\rangle_{B}\nonumber\\
&&+|0\rangle_{A}|2,a\rangle_{B}+|0\rangle_{A}|2,b\rangle_{B} \biggr].
\end{eqnarray}
Here, $|0\rangle$ is the empty state and for the second line, $|\cdot\rangle_{A(B)}$ is a single-particle state in A(B)-subsystems.
From this state $|W_{-,1}\rangle$, the density matrix is given by $\rho_{AB}=|W_{-,1}\rangle \langle W_{-,1}|$. 
Moreover, by tracing out the B-subsystem we can obtain the partial density matrix directly: 
\begin{eqnarray}
\rho_{A}&=&\frac{1}{4}\biggl[ |1,a\rangle_{A}\langle 1,a|_{A}+i |1,a\rangle_{A}\langle 1,b|_{A}\nonumber\\
&&-i|1,b\rangle_{A}\langle 1,a|_{A} +|1,b\rangle_{A}\langle 1,b|_{A}+2|0\rangle_{A}\langle 0|_{A}\biggr].\nonumber\\
\end{eqnarray}
The $\rho_{A}$ has eigenvalues,  $\lambda_{1}=1/2$,$\lambda_{2}=1/2$, and $\lambda_{3}=0$. 
Accordingly, the entanglement entropy for the two-site A-subsystem is $s_{A}=-\sum_{\ell}\lambda_{\ell}\log \lambda_{\ell}=\log 2$.
Therefore, when we cut the single lower CLS in half we get $\log 2$ as the basic unit of the contribution of entanglement entropy. 
Actually, the basic unit of the contribution of the entanglement entropy is analogs to the contribution from a single singlet state on the strong link in the Su-Sherifer-Heeger model \cite{Ryu}. 
In addition, even for a single upper CLS, $W^{\dagger}_{+,1}|0\rangle$, we also obtain $\log 2$ as the contribution of entanglement entropy in system. 
This basic contribution also becomes the signal of the presence of an edge mode.  
If at half-filling the groundstate is given by $|{\rm GS1}\rangle$ of Eq.~(7) in the main text, 
then when we cut the system into two helves, 
the subsystem entanglement entropy can only be obtained by cutting only two lower CLSs located on each boundaries. 
Hence, the subsystem entanglement entropy of $|{\rm GS1}\rangle$ becomes $2\log 2$. 
This contribution implies the presence of an edge mode as a non-trivial topological phase \cite{Ryu,Bermudez2}.
For $|{\rm {\rm GS1}}\rangle$, at any sub-system size, we get $2\log 2$ as the subsystem entanglement entropy. 

\section*{Appendic C: Entanglement entropy for a single pair CLS on four site system}
\begin{figure}[h]
\begin{center} 
\includegraphics[width=3.5cm]{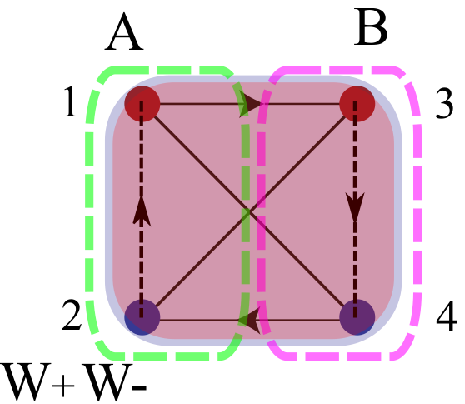}
\end{center} 
\caption{Single pairing state of the lower and upper flat-band CLSs}
\label{FigA2}
\end{figure}
For $\Delta_{0}>2t_0$, the system exhibits the groundstate $|{\rm GS2}\rangle$. 
Here, for odd lattice site $j$, both lower and upper CLSs are located, i.e., $W^{\dagger}_{j,+}W^{\dagger}_{j,-}|0 \rangle$, 
which can be regarded as a pairing of the CLSs.  
At a grace, when the single pairing state is cut into two helves, the entanglement entropy may be two times of $\log 2$ since the two CLSs are cut. 
This intuitive picture is however not correct as demonstrated in detail below. 
As already done for the single CLS on a single plaquette, we calculate the entanglement entropy for the single pairing state on a single plaquette by using Peschel's formulation \cite{Peschel,Ryu}.
First, we calculate the subsystem correlation function:     
\begin{eqnarray}
C^{corr}=\langle c^{\dagger}_{i}c_{j}\rangle &=&|W_{1,-}\rangle \langle W_{1,-}|+|W_{1,+}\rangle \langle W_{1,+}|\nonumber\\
           &=& \left[
\begin{array}{rrrr}
  1/2  & i/2   & 0 & 0 \\
  -i/2  & 1/2  & 0 & 0 \\
   0     &    0 & 1/2  & -i/2 \\
   0     &   0  &  i/2 & 1/2 
\end{array}
\right],
\end{eqnarray}
where $i,j=1,2,3,4$ as shown in Fig.~\ref{FigA2}, and we used $|W_{1,+}\rangle =(1/2)[i,1,1,i]^{t}$, $|W_{1,-}\rangle =(1/2)[i,1,-1,-i]^{t}$. 
Second, we truncate the correlation matrix into two halves (we call the either one the A-subsystem):
\begin{eqnarray}
C^{corr}_{i,j\in A}=
\left[
\begin{array}{rr}
  1/2  & i/2   \\
  -i/2  & 1/2 
\end{array}
\right],
\label{2CLScorr_matrix}
\end{eqnarray}
The eigenvalues of the above matrix are either zero or one, i.e, the subsystem exhibits no maximally entanglement mode (MEM) \cite{Pouranvari}.
Therefore, all eigenmodes do not contribute the entanglement entropy of the A-subsystem. 

\section*{Appendix D: Calculation of Zak phase from real space representation}
The Zak phase is calculated from the slater determinant given by single particle eigenstates in real space representation.
We introduce the twisted boundary phase condition. 
The hopping parameters on boundary are changed: $t_{0}\to t_{0}e^{i\theta}$, $t_{1}\to t_{1}e^{i\theta}$, where $\theta \in (0,2\pi]$. 
In practical calculation we employ a discretized $\theta_{\ell}=2\pi \ell/N_{\theta}$ where $\ell=1,2,\cdots, N_{\theta}$.
The many particle wave function represented by real space and characterized by $\theta_{\ell}$ is given by using single particle states,
\begin{eqnarray}
|\Psi (\theta_{\ell})\rangle = \frac{1}{\sqrt{N_{F}!}}
\begin{vmatrix}
|\psi^{1}_{\theta_{\ell}}\rangle & \cdots &|\psi^{N_{F}}_{\theta_{\ell}}\rangle\\
\vdots & \ddots  & \vdots \\
|\psi^{1}_{\theta_{\ell}}\rangle & \cdots &|\psi^{N_{F}}_{\theta_{\ell}}\rangle
\end{vmatrix}
\label{WF}
\end{eqnarray}
where $|\psi^{\alpha}_{{\theta}_{\ell}}\rangle$ is $\alpha$-th single particle state, $\alpha=1,\cdots, N_{F}$, 
and $|\psi^{\alpha}_{{\theta}_{\ell}}\rangle$ is represented by real space bases.
$N_{F}$ is the number of maximum occupied state, determined by Fermi energy. 
Here, by using $|\Psi(\theta_{\ell})\rangle$ the Zak phase is given by a discretization form \cite{Vanderbilt}
\begin{eqnarray}
\phi_{Zak}=-\sum^{N_{\theta}}_{\ell=1}{\rm Im} \ln \langle \Psi(\theta_{\ell})|\Psi(\theta_{\ell+1})\rangle, 
\label{Zak}
\end{eqnarray}
where the inner product is given by the single particle eigenstates,
\begin{eqnarray}
\langle \Psi(\theta_{\ell})|\Psi(\theta_{\ell+1})\rangle =
\begin{vmatrix}
\langle \psi^{1}_{{\theta}_{\ell}}|\psi^{1}_{{\theta}_{\ell+1}}\rangle & \cdots &\langle \psi^{1}_{{\theta}_{\ell}}|\psi^{N_{F}}_{{\theta}_{\ell+1}}\rangle\\
\vdots & \ddots  & \vdots \\
\langle \psi^{N_{F}}_{{\theta }_{\ell}}|\psi^{1}_{{\theta}_{\ell+1}}\rangle & \cdots & \langle \psi^{N_{F}}_{{\theta}_{\ell}}|\psi^{N_{F}}_{{\theta}_{\ell+1}}\rangle
\end{vmatrix}.\nonumber
\label{Cmatrix}
\end{eqnarray}
The Zak phase of Eq.~(\ref{Zak}) is actually gauge invariant. Although an arbitrary global phase are attached to each eigenstate when it is generated numerically, 
we can obtain correct values of the Zak phase.
Of course, the Zak phase of Eq.~(\ref{Zak}) can be also calculated from the momentum eigenstate obtained by the bulk momentum Hamiltonian $H_{\rm Cd}(k)$ of  Eq.~(6) in the main text. 
Each band results of the Zak phase gives same results with the results obtained by using Eq.~(\ref{Zak}).


\end{document}